\documentclass[a4paper,11pt]{article}
\usepackage{pos,subcaption}

\title{Axions explain the formation of supermassive \\
black holes at cosmic dawn} 

\author*[a]{Pierre Sikivie}
\author[a]{Yuxin Zhao}
\affiliation[a]{Department of Physics, University of Florida,\\ Gainesville, Florida 32611, USA}
\emailAdd{sikivie@ufl.edu}

\date{January 2, 2026}

\abstract{
In a recent paper \cite{smbh} we pointed out that supermassive 
black holes, with masses ranging from $10^5$ to $10^{10} M_\odot$ 
form naturally at cosmic dawn if the dark matter is QCD axions or 
axion-like particles with mass $m > 10^{-16}$ eV/$c^2$.  No additional  
assumptions are required.  Here we answer in detail the most 
commonly raised questions regarding our work.
}

\FullConference{3rd General Meeting of the COST Action COSMIC WISPers (CA21106) - 3rd Training School of the COST Action COSMIC WISPers (CA21106) (COSMICWISPers2025)\\
9–12 Sept 2025 and 16–19 Sept 2025\\
Sofia, Bulgaria and Annecy, France\\}

\begin{document}

\maketitle

In 1963  Maarten Schmidt published the discovery of an extraordinarily 
luminous point-like source at redshift z = 0.158, 3C273, the first known 
quasar \cite{MS}.  Many similar objects were discovered in the following 
years.   They are collectively named `active galactic nuclei' or AGN for 
short.  AGN are very powerful emitters with luminosity of order $10^{13} 
L_\odot$.  Yet they are variable on time scales as short as a few hours,
implying that they are of relatively small spatial extent.  After a 
while, it was recognized that AGN are powered by accretion onto 
supermassive black holes \cite{Rees}. Furthermore, it became clear
that most galaxies have a supermassive black hole at their center, 
with  mass ranging from $10^5$ to $10^{10} M_\odot$ \cite{Korm}.  
The event horizons of the black holes at the center of the Milky Way 
and at the center of the large galaxy M87 have been imaged using
very long baseline interferometry \cite{EHT}.  Recently, observations
using the James Webb Space Telescope have shown that powerful AGN
are already present at cosmic dawn, i.e. as early as $z \sim 10$ 
\cite{JWST}.

How the supermassive black holes form has been an enduring 
mystery. The main impediment to their formation is conservation 
of angular momentum. If, for example, a $10^8 M_\odot$ black hole 
condenses out of a region of density $10^{-25}$ gr/cc, which
is a value typical of the cosmological energy density at cosmic 
dawn, the material forming the black hole must shrink in all 
directions by nine orders of magnitude. Even a very small 
amount of angular momentum keeps infalling material from 
entering the black hole by introducing a distance of closest 
approach.  Angular momentum can be transported outward 
if the material has viscosity but in that case the material 
heats up and acquires pressure opposing its compression. 
See Ref. \cite{bhfrev} for a review of the issues involved 
in supermassive black hole formation using conventional 
physics.  

To overcome the difficulties arising when only 
conventional physics is involved, it has been 
proposed that the supermassive black holes form 
as a result of the gravothermal collapse of 
overdensities of dark matter with very strong
self-interactions \cite{SS} or by accretion
onto dark stars, i.e. stars that are powered by dark
matter annihilation \cite{KF}. In such scenarios,
seed black holes form that have masses of order
$10^5 \text{--} 10^6~M_\odot$. The larger supermassive
black holes with mass $10^7 \text{--} 10^{10}~M_\odot$
are then supposed to be the result of accretion
onto, and mergers of, the seed black holes.
Such proposals are severely challenged by the 
discovery of powerful AGN at cosmic dawn because 
there is too little time to grow the black holes 
powering them.  Ref. \cite{Qin} indicates how the 
amount of available time may be increased somewhat 
by modifying the spectrum of primordial density 
perturbations.  One may contemplate  the possibility 
that the supermassive black holes are primordial in 
nature, i.e. that they formed long before cosmic 
dawn, but this proposal tends to run afoul of 
constraints from cosmic microwave background 
observations.

In ref. \cite{smbh} we showed that supermassive
black holes form naturally at cosmic dawn if the
dark matter is QCD axions or axion-like particles
with mass $m > 10^{-16}$ eV. No additional assumptions 
are required. The crucial step is to recognize that 
cold dark matter axions thermalize by their 
gravitational self-interactions \cite{CABEC,Erken}. 
When the axions thermalize they form a Bose--Einstein 
condensate, meaning that most axions go to the lowest 
energy state available to them through the thermalizing
interactions. When an axion overdensity collapses
near cosmic dawn, the gravitational self-interactions
among the axions produce a long range viscosity
that causes outward transport of angular momentum.
The heat produced in the axion case flows into
the thermal distribution accompanying the condensate
whereas the condensate, containing most of the
axions, stays in the lowest energy particle state
available. That state is one of rigid rotation
where most of the angular momentum resides far
from the central overdensity which may then
collapse into a black hole.

In our proposal, the supermassive black hole
forms at the precise moment of initial collapse 
of an overdensity, the thermalization and 
transport of angular momentum having taken place
before that moment.  A galaxy forms afterwards 
by infall around the already formed supermassive 
black hole.  Our proposal is consistent with the 
large ratio, $M_{\rm BH}/M_* > 2$, of black 
hole mass $M_{\rm BH}$ to stellar mass $M_*$ 
observed in the early AGN, called Little Red 
Dots, observed near cosmic dawn \cite{ratio}.

In the refereeing process, and in discussions
and talks, a number of questions were raised
in regard to the supermassive black hole formation
mechanism proposed in ref. \cite{smbh}.  Below 
is a list of such questions and the answers we 
give.  A thorough reading of ref. \cite{smbh} 
is a prerequisite. 

\vskip 0.5cm

{\bf Shouldn't the velocity dispersion be identified with the virial 
velocity of the collapsed halo?}

\vskip 0.2cm

No, our paper discusses the thermalization of the axion fluid 
before the initial collapse of an axion overdensity, i.e. at 
$t < t_c$ where $t_c$ is the collapse time.  Our paper does 
not discuss the thermalization of a virialized or virializing 
halo formed after an axion overdensity has collapsed, i.e. at 
$t > t_c$.

The relevant velocity dispersion for thermalization by gravitational 
self-interactions of a collapsed halo is indeed its virial velocity.
The corresponding thermalization rate is extremely small and 
ineffective for the purpose of supermassive black hole formation.  

\vskip 0.5cm

{\bf What then is the relevant velocity dispersion?}

\vskip 0.2cm

Fig. 1, reproduced from ref. \cite{smbh}, shows where the axions are 
located in phase space at four different times up to the collapse time
$t_c$.  Our paper discusses the thermalization of the axion fluid 
within the thin layers of the 3-dimensional sheet on which the axions 
lie in phase space before and up to the collapse time.  The relevant 
velocity dispersion is the thickness of that phase-space sheet, i.e. 
the thickness of the lines in Fig. 1.

Appendix A of ref. \cite{smbh} discusses the numerous particle 
states of the axions within the phase space sheet, and shows that 
the interference between those particle states produces large density 
fluctuations and hence large gravitational field fluctuations which 
rearrange the distribution of the axions among the particle states 
within the phase space sheet, i.e.  that the axion gravitational 
self-interactions produce a thermal relaxation of the axions within 
the phase space sheet.   The appendix shows that the axion fluid has 
density fluctuations 
$\delta \rho = \rho$ and that these fluctuations are correlated over 
distances 
\begin{equation}
\ell \simeq {1 \over m \delta v}
\label{ell}
\end{equation}
where $m$ is the axion mass and $\delta v$ is the velocity dispersion.
$\ell$ is called the correlation length.

\vskip 0.5cm

{\bf What then is the correlation length of the cold axion fluid?}

\vskip 0.2cm 

$\ell$ is of order the horizon size when the axion mass turns on during 
the QCD phase transition. After that phase transition, $\ell$  stretches 
with the expansion of the Universe. When the photon temperature is of 
order keV and the Universe is of order one month old, the axion fluid 
thermalizes by gravitational self-interactions for the first time 
\cite{CABEC,Erken}. Bose--Einstein condensates form and grow in size 
until they are of order the horizon scale at the time.  Within a 
Bose--Einstein condensate the correlation length is the size  of the 
Bose--Einstein condensate \cite{Chak}.  After the initial period 
of thermalization near keV photon temperature, $\ell$ stretches again 
with the Universe’s expansion. At cosmic dawn, it is of order Mpc, 
large enough to contain the collapsing axion overdensities that 
we discuss.

\vskip 0.5cm

{\bf Shouldn't the relaxation rate by gravitational 
self-interactions be proportional to the square of 
Newton's gravitational constant?}

\vskip 0.2cm

No, not in the situation under consideration.  In the situation 
under consideration the correct formula for the relaxation rate 
by gravitational self-interactions is of order $G$, not of order $G^2$.

The most familiar formula for the relaxation rate of a fluid of 
interacting particles is
\begin{equation}
\Gamma \sim n~\sigma~\delta v
\label{famil}
\end{equation}
where $n$ is the number density of particles, $\sigma$ is their scattering 
cross-section and $\delta v$ their velocity dispersion.  For relaxation by 
gravitational scattering, the cross-section is of order
\begin{equation}
\sigma_g \sim {(2 m G)^2 \over (\delta v)^4}~~\ .
\label{gs}
\end{equation}
Eqs.~(\ref{famil}) and (\ref{gs}) may be used to estimate the relaxation 
rate by gravitational scattering of axions in a collapsed halo.  $\delta v$
is the halo virial velocity in that case.  The relaxation rate will
be found to be extremely small and negligible compared to the dynamical 
evolution rate $\sqrt{G n m}$.

Eq.~(\ref{famil}) is not applicable to the situation under 
consideration because the axion fluid is a degenerate Bose gas.
For a degenerate Bose gas,  Eq.~(\ref{famil}) is modified to
\begin{equation}
\Gamma_k \sim n~\sigma_g~\delta v~({\cal N} + 1)
\label{gk}
\end{equation}
where ${\cal N}$ is the quantum degeneracy of the particles, 
i.e. the average occupation number of the states that the 
particles occupy, or, equivalently, their phase-space density 
in units of $(2 \pi \hbar)^{-3}$.  Be aware however that 
Eq.~(\ref{gk}) assumes that  $\Gamma_k < \delta E$ where 
$\delta E$ is the energy dispersion of the particles.  The 
condition $\Gamma_k < \delta E$ is necessary for the 
validity of Fermi's Golden Rule on which Eqs.~(\ref{famil})
and (\ref{gk}) are based.   To illustrate this, let us 
apply Eq.~(\ref{gk}) to the collapsing axion overdensity 
and see what happens.

At $t_{\rm in} \simeq$ 240 Myr, when the central 
part of an axion overdensity that collapses
at redshift $z_c \simeq 10$ is at turnaround,
i.e. at the moment it stops expanding for the 
first time and begins to contract, the average 
cosmological energy density is 
$\bar{\rho} \simeq 1.3 \cdot 10^{-26}$ gr/cc.
Hence the average cosmological axion number 
density 
\begin{equation}
n = {\bar{\rho} \over m}
\simeq 0.8 \cdot 10^{13}/{\rm cm}^3
\left({\mu{\rm eV} \over m}\right)~~\ .
\label{den}
\end{equation}
Inside an overdensity $n$ is somewhat larger 
than its average cosmological value but let us 
ignore  this relatively unimportant distinction.

The axions that scatter off each other are  
in modes of wavelength of order the size $r$  
of the overdensity.  Hence
\begin{equation}
\delta v \sim {\hbar \over m r}
= 2 \cdot 10^{-21} c \left({\mu{\rm eV} \over m}\right)
\label{delv}
\end{equation}
where we took $r \sim 10^{22}$ cm as in Eq.~(5)
of ref. \cite{smbh}. Therefore
\begin{equation}
{\cal N} \sim  n
{(2 \pi \hbar)^3 \over {4 \pi \over 3} (m \delta v)^3}
\sim 2.5 \cdot 10^{81}
\left({\mu{\rm eV} \over m}\right)
\label{deg}
\end{equation}
and
\begin{equation}
\sigma_g \simeq 0.5 \cdot 10^{-50} {\rm cm}^2
\left({m \over \mu{\rm eV}}\right)^6~~\ .
\label{cx}
\end{equation}
$\sigma_g$ is very small but ${\cal N}$ is  
very large.  Using Eq.~(\ref{gk}) we have
\begin{equation}
\Gamma_k \sim 3 \cdot 10^{34} {1 \over {\rm sec}}
\left({m \over \mu{\rm eV}}\right)^3~~\ .
\label{gkr}
\end{equation}
A relaxation rate as large as  
$~3 \cdot 10^{34}~{\rm sec}^{-1} \simeq 
2 \cdot 10^{10}~{\rm GeV}/\hbar$ cannot possibly 
be correct for reasons that we now make clear.

Eq.~(\ref{gk}) assumes that Fermi's
Golden Rule can be applied to the individual
$a + a \rightarrow a + a$ scatterings.  This is
valid only if each scattering happens separately 
from all the other scatterings so that energy
is conserved in each scattering separately.  This
assumption fails here because the scatterings happen
so fast that they overlap in time.  In $10^{-34}$
seconds, one can only distinguish particle states that
differ in energy by more than $10^{34}~(\rm sec)^{-1}
\simeq 6.7~10^{9}$ GeV.  The energy dispersion of
the particles in the fluid is much less than that.

The relevant condition for Eq.~(\ref{gk}) to be 
applicable is that the  energy dispersion $\delta E$
of the particles in the fluid is much larger than
the relaxation rate $\Gamma$.  When $\delta E \gg
\Gamma$, in the so-called particle kinetic regime,
the standard formula Eq.~(\ref{gk}) applies.
When $\delta E \ll \Gamma$, in the so-called condensed
regime, the correct formula is Eq. (3) of our
paper, i.e.
\begin{equation}
\Gamma_c \sim 4 \pi G \rho m \ell^2 {1 \over \hbar}
\label{gc}
\end{equation}
where $\rho = n m$ and $\ell = \hbar/ m \delta v$.
The condensed regime formula is derived, and discussed
in detail, in ref. \cite{Erken}.  The formula is not 
so familiar because very few systems thermalize in the 
condensed regime.  The cold axion fluid is in fact 
the only such system that we are aware of. In  
 ref. \cite{smbh} we gave the following intuitive
justification for the formula.  Since the axion 
fluid has density fluctuations $\delta \rho = \rho$ 
correlated over distances of order $\ell$, the 
gravitational field has fluctuations of order 
$\delta g \sim 4 \pi n m \ell$.  The gravitational 
field fluctuations  change all the particle momenta 
by order the momentum dispersion 
$\delta p =m  \delta v \sim \hbar/\ell$ in a 
characteristic time  $\tau \sim \delta p/m \delta g$.  
$\Gamma_c$ is the inverse of that time.

\vskip 0.5cm 

{\bf Can the axion fluid be described by the 
Schr\"odinger--Poisson equations?}

\vskip 0.2cm

The phenomenon that we discuss cannot be described by the 
Schr\"odinger--Poisson equations. The Schr\"odinger--Poisson 
equations are the classical equations of motion of the
self-gravitating axion field.  The phenomenon we describe 
occurs only in the quantum field description of the axion 
field.   When using the Schr\"odinger--Poisson equations, one 
is not doing quantum mechanics. For one thing, quantum mechanics 
is linear whereas the Schr\"odinger--Poisson equations are 
non-linear.  The Schr\"odinger--Poisson equations are to the 
quantum field theory of self-gravitating axions what Maxwell’s 
equations are to quantum electrodynamics.

Einstein pointed out that when a highly degenerate Bose 
fluid thermalizes and the number of particles is conserved, 
a fraction of order one of the particles go to the lowest 
energy state available to them through the thermalizing 
interactions. Bose--Einstein condensation plays an essential 
role in the process we describe because the lowest energy state 
for given angular momentum is one of rigid rotation where most 
of the angular momentum has moved towards the periphery.  If we 
were to describe the axion fluid by the Schr\"odinger--Poisson 
equations, there would be no Bose--Einstein condensation. 
The result of thermalization would instead be energy 
equipartition among all the axion field modes as this is 
the outcome of thermalization in all classical field 
descriptions. When every field mode has the same energy, 
an ultraviolet catastrophe ensues first pointed out by 
Raleigh in the context of electrodynamics. A second reason 
why the phenomenon we discuss  cannot be captured by the 
Schr\"odinger--Poisson equations is that these equations 
forbid the generation of vorticity. Rigid rotation requires 
vorticity.

\vskip 0.5cm

{\bf Isn't it so that the classical and quantum descriptions 
of field oscillators agree with one another to order $1/{\cal N}$?}

\vskip 0.2cm

No, it is not so.

The quantum and classical descriptions of field oscillators
agree to order $1/{\cal N}$ when they are non-interacting. 
But the system under consideration has gravitational 
self-interactions. In an interacting system the deviations 
of the classical description from the quantum description 
are of order $1/{\cal N}$  only on time scales that are short 
compared to the thermal relaxation timescale. On time 
scales large compared to the thermal relaxation timescale, 
the classical and quantum systems behave completely differently.  
In the  collapsing axion fluid, the thermal relaxation time is 
of order 1 second.

\vskip 0.5cm

{\bf Can the initial collapse of an axion overdensity result in 
a Bose star?}

\vskip 0.2cm

No Bose star forms in the initial, i.e. at time $t_c$, 
collapse of an axion overdensity assuming the axion mass 
is larger than $10^{-16}$ eV/$c^2$.  A Bose star may form 
later in the post-collapse virializing halo, but we do 
not discuss this.

The conventional description of a Bose star makes two assumptions: 
1) that all Bosons are in the same particle state described 
by a wavefunction, and 2) that the star is static or slowly 
evolving. The wavefunction is then obtained by solving the 
Schr\"odinger--Poisson equations. We make neither of those 
two assumptions. We do not make assumption 1) because in our 
description of the collapsing axion fluid the axions occupy 
many different particle states, each with its own wavefunction. 
The wavefunctions differ from each other by long wavelength 
modulations as described in Appendix A of ref. \cite{smbh}. 
Thermal relaxation means that the axions move between those 
many different highly occupied particle states, increasing 
the system entropy.  We do not make assumption 2) because 
the axion fluid is collapsing. It is neither static nor 
quasi-static.

In a Bose star, the gravitational attraction is balanced 
by “quantum pressure” (The name quantum pressure is unfortunate 
because it suggests that the pressure is a property of the 
quantum field theory whereas it is in fact a property of the 
classical field theory.) Because of the balance between 
gravitational attraction and quantum pressure, the Bose star 
is in a static or quasi-stationary state. In the collapsing 
axion fluid that we discuss nothing balances the gravitational 
attraction as the quantum pressure is negligible at all 
times assuming as we do that the axion mass is larger 
than $10^{-16}$ eV/$c^2$.
 
\begin{acknowledgments}

We thank Wenzer Qin and Qiushi Wei for useful discussions.
This article is based upon work from COST Action COSMIC WISPers CA21106, supported by COST 
(European Cooperation in Science and Technology).

\end{acknowledgments}

\begin{figure}
\begin{center}
\includegraphics[height=110mm]{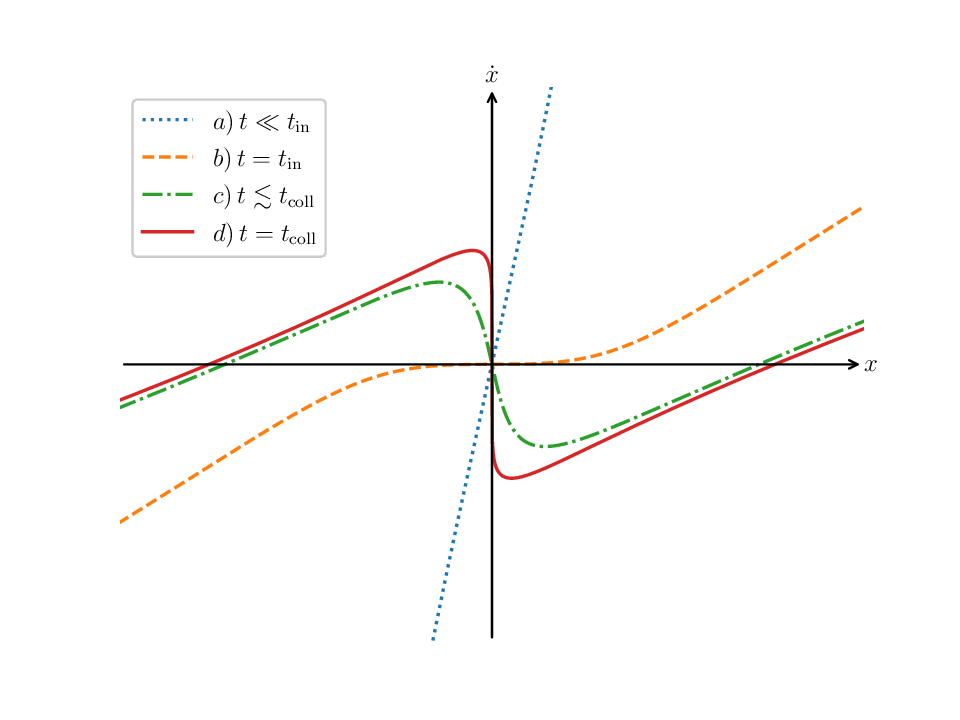}
\vspace{0.3in}
\caption{Phase space distribution of cold collisionless
particles during the collapse of a large smooth overdensity 
near cosmic dawn, at four different times: a) just after 
the Big Bang, b) when the central part of the overdensity 
is at turnaround, c) just before, and d) at the time 
$t_c$ of collapse of the central overdensity. 
$x$ is the spatial coordinate along an arbitrary direction 
through the overdensity. An actual overdensity has small 
scale structure which has been smoothed out in the figure.}
\end{center}
\label{fig:phsp}
\end{figure}

\clearpage

\end{document}